\documentclass{svjour2}
\smartqed
\usepackage{graphicx}
\usepackage{amsfonts}
\usepackage{amsmath,amssymb,bbm}
\usepackage{verbatim}
\usepackage{amsmath}
\usepackage{amssymb}
\usepackage{amsfonts}
\usepackage{latexsym}
\newcommand {\nn} {\nonumber}

\newcommand {\bd}{\begin{document}}
\newcommand {\ed}{\end{document}}
\newcommand {\be}{\begin{equation}}
\newcommand {\ee}{\end{equation}}
\newcommand {\ba}{\begin{eqnarray}}
\newcommand {\ea}{\end{eqnarray}}
\newcommand {\bc}{\begin{center}}
\newcommand {\ec}{\end{center}}
\newcommand {\ul}{\underline}
\newcommand {\tc} {\textcolor}
\newcommand{\comm}[2]{\left[ #1 , #2 \right]}
\begin{document}
\title{Causality and Statistics on the Groenewold - Moyal Plane}
\author{A. P. Balachandran$^{\star \dagger}$ \and Anosh Joseph \and Pramod Padmanabhan}
\institute{A. P. Balachandran \at
              Department of Physics, Syracuse University, Syracuse, NY 13244-1130, USA \\
          Departamento de Mathem\'aticas, Universidad Carlos III de Madrid, 28911 Legan\'es, Madrid, Spain\\
              \email{bal@phy.syr.edu}
           \and
           Anosh Joseph \at
              Department of Physics, Syracuse University, Syracuse, NY 13244-1130, USA \\
              \email{ajoseph@phy.syr.edu}
              \and
           Pramod Padmanabhan \at
            Department of Physics, Syracuse University, Syracuse, NY 13244-1130,
            USA \\ \email{ppadmana@syr.edu}
       \and
$^{\star}$ Based on the talk given by APB at the Workshop {\it Theoretical and Experimental Aspects of the Spin Statisics Connection and Related Symmetries}, Stazione Marittima Conference Center, Trieste, Italy from the 21st to the 25th of October 2008.\\$^{\dagger}$ C\'atedra de Excelencia\\PREPRINT NO. SU-4252-892}
\date{Received: date / Accepted: date}
\maketitle
\begin{abstract}
Quantum theories constructed on the noncommutative spacetime called the Groenewold-Moyal plane exhibit many interesting properties such as Lorentz and CPT noninvariance, causality violation and twisted statistics. We show that such violations lead to many striking features that may be tested experimentally. These theories predict Pauli forbidden transitions due to twisted statistics, anisotropies in the cosmic microwave background radiation due to correlations of observables in spacelike regions and Lorentz and CPT violations in scattering amplitudes.
\end{abstract}
\section{Introduction}
\label{intro}
The connection between spin and statistics is established in local quantum field theories by the requirement of causality \cite{Bogoliubov}, \cite{Haag}, \cite{Weinberg}. The condition of locality is generally expressed in such theoretical frameworks by the assumption that the observables of spacelike separated observers commute.

In the theory of response functions in physical systems, the Kramers-Kronig relations connect the real and imaginary parts of the response function by making use of the fact that causality implies analyticity and vice versa.

A physical system should not respond before the time at which it is disturbed. If $R(t)$ is the response and disturbance of the system is zero for time $t<0$,
\begin{equation}
R(t)=0,~~~~t<0,
\end{equation}
then its Fourier transform
\begin{equation}
\widetilde{R}(\omega)=\int_{-\infty}^\infty dt e^{i\omega t} R(t) = \int_0^\infty dt e^{i\omega t} R(t)
\end{equation}
being holomorphic in
\begin{equation}
\textrm{Im}~\omega > 0
\end{equation}
leads to Kramers-Kronig relations.

Causal set theory, a discrete and Lorentz invariant approach to quantum gravity, rests on the central hypothesis that spacetime {\it is} a causal set \cite{Henson}, \cite{Dowker}, \cite{Sorkin}. In a causal set $C$, the binary (partial order) condition $\succ$ between its two elements $x$ and $y$ reads:
\begin{center}
``$x \succ y$, if $x$ is to future of $y$."
\end{center}
\begin{figure*}
\includegraphics[width=6cm]{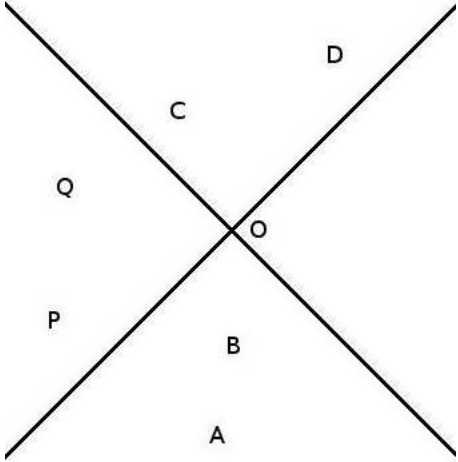}
\caption{The observer is at the spacetime point $O$. Events $C$, $D$ are in the future, and $A$, $B$ are in the past light cone of $O$. They are causally related to $O$. Events $Q$, $P$ are spacelike relative to $O$ and are not causally related to $O$.}
\label{fig:1}
\end{figure*}

In local quantum field theories, two observables $\rho (x)$, $\eta (y)$ commute if $x$ and $y$ are spacelike separated:
\begin{equation}
\comm{\rho(x)}{\eta(y)}_-=0
\end{equation}
if
\begin{equation}
(x^0-y^0)^2-(\vec{x}-\vec{y})^2<0~~~\textrm{or}~~~x\sim y.
\end{equation}

For scalar fields the above relation takes the form of a commutator
\begin{equation}
\comm{\varphi(x)}{\chi(y)}_-=0~~~x\sim y,
\end{equation}
and for spinor fields it takes the form an anti-commutator
\begin{equation}
\comm{\psi^{(1)}_\alpha(x)}{\psi^{(2)}_\beta(y)}_+=0~~~x\sim y.
\end{equation}
These relations also express statistics of the quantum fields. So causality and statistics are connected.

It is interesting to study how the connection between causality and statistics is affected in quantum field theories that exhibit features like nonlocality, Lorentz noninvariance etc. A quantum field theory based on the noncommutativity of spacetime shows these interesting features. There are indications both from theories of quantum gravity and string theory that spacetime is noncommutative with length scale of the order of Planck length. We can model such spacetime noncommutativity using the algebra of functions called the Groenewold-Moyal (GM) plane.

The GM plane describes noncommutative spacetime where commutation relations and hence causality and statistics are deformed.

\section{The Groenewold-Moyal Plane}
The GM plane is the algebra $\mathcal A_\theta$ of smooth functions on $\mathbb{R}^{d+1}$ with a twisted product:
\be
f \star g (x) = f(x)~e^{\frac{i}{2}{\overleftarrow{\partial}}_{x^{\mu}} \theta ^{\mu \nu}
{\overrightarrow{\partial}} _{y^{\nu}}}~g(y)\Big|_{x=y},
\ee
where $\theta ^{\mu \nu} = -\theta ^{\nu \mu} =\textrm{constant}$.

It implies the commutation relation
\be
(\widehat{x}_{\mu} \star \widehat{x}_{\nu} - \widehat{x}_{\nu} \star \widehat{x}_{\mu}) = \left [ \widehat{x}_\mu, \widehat{x}_\nu \right ]_{\star} = i \theta_{\mu \nu},~~\mu, \nu = 0, 1, \cdots, d,
\ee
with
\be
\widehat{x}_{\mu} = \textrm{coordinate functions},~~~~\widehat{x}_{\mu}(x) = x_{\mu}.\nn
\ee

We will describe a particular approach to the formulation of quantum field theories on the GM plane and indicate its  physical consequences.

It is interesting to see how noncommutative structure of spacetime emerges at very small length scales from the arguments based on Heisenberg's uncertainty principle and Einstein's theory of classical gravity. Doplicher, Fredenhagen and Roberts \cite{Doplicher} give the following arguments

In order to probe physics at the Planck scale $L$, the Compton wavelength $\hbar/Mc$ of the probe must fulfill
\begin{equation}
\frac{\hbar}{Mc}~\leq~L~~~\textrm{or}~~~M~\geq~\frac{\hbar}{Lc}~\simeq~\textrm{Planck mass}.\nn
\end{equation}

Such high mass in the small volume $L^3$ will strongly affect gravity and can cause black holes and their horizons to form. This suggests a fundamental length limiting spatial localization indicating {\it space-space noncommutativity}.

Similar arguments can be made about {\it time-space noncommutativity}. Observation of very short time scales requires very high energies. They can produce black holes and black hole horizons will then limit spatial resolution suggesting
\begin{equation}
\Delta t~\Delta |{\overrightarrow x}|~\geq\ L^2~,~~~L=\textrm{a fundamental length.}\nn
\end{equation}

The GM plane {\it models} above spacetime uncertainties.
\section{The Twisted Coproduct}
If there is a symmetry group $G$ with elements $g$ and it acts on single particle Hilbert spaces ${{\cal H}_i}$ by unitary representations $g \rightarrow U{_i}(g)$, then conventionally it acts on ${\cal H}_1 \otimes {\cal H}_2$ by the representation
\begin{equation}
g \rightarrow [U_1 \otimes U_2]( g \times g).
\label{twoparticle}
\end{equation}

The homomorphism
\begin{eqnarray}
\Delta: G &\rightarrow& G \times G, \nn \\
g &\rightarrow& \Delta(g) := g \times g \nn
\end{eqnarray}
underlying these equations is said to be a coproduct on $G$.

The action of $G$ on multiparticle states involves more than just group theory. It involves the coproduct $\Delta$.

The $\star$-multiplication between two functions $f$ and $g$ on the noncommutative algebra can be expressed in terms of the {\it twist element} \cite{s-matrix}, \cite{Drinfel'd1}, \cite{Drinfel'd2}, \cite{Chaichian}, \cite{Chaichian1}
\begin{equation}
{\cal F}_\theta = e^{\frac{i}{2} \partial_\mu \otimes \theta^{\mu \nu}\partial_\nu}
\end{equation}
as follows
\begin{equation}
f \star g = m_0 \cdot {\cal F}_ \theta (f \otimes g),
\end{equation}
where $m_0$ is the point-wise multiplication map of the commutative algebra $\mathcal A_0$:
\begin{equation}
m_{0} (\alpha \otimes \beta)(x) = \alpha(x)\beta(x)\\
\end{equation}

Let $\Lambda$ be an element of the connected component of the Poincar\'{e} group ${\cal P}_+^\uparrow$. Then for $x \in {\mathbb R}^N$ we have
\begin{equation}
\Lambda: x \rightarrow \Lambda x \in {\mathbb R}^N.
\end{equation}
It acts on functions on ${\mathbb R}^N$ by pull-back:
\begin{equation}
\Lambda: \alpha \rightarrow \Lambda^* \alpha, \quad (\Lambda^* \alpha)(x) =
\alpha[\Lambda^{-1}x].
\end{equation}

The work of Aschieri {\it et al.} \cite{Aschieri} and Chaichian {\it et al.} \cite{Chaichian} based on Drinfel'd's original work \cite{Drinfel'd1}, \cite{Drinfel'd2} shows that ${\cal P}_+^\uparrow$ acts on ${\cal
A}_\theta ({\mathbb R}^N)$ compatibly with $m_\theta$ if its coproduct is ``twisted'' to $\Delta_\theta$ where
\begin{equation}
\Delta_\theta (\Lambda) = {\cal F}_\theta^{-1} (\Lambda \otimes \Lambda) {\cal F}_\theta.
\end{equation}
\section{The Twisted Statistics}
The action of the twisted coproduct is not compatible with standard statisitcs. Statistics also should be twisted in quantum theory.

A two-particle system for commutative case ($\theta^{\mu \nu}=0$) is a function of two sets variables and it lives in $\mathcal A_0 \otimes \mathcal A_0$. It transforms according to the usual coproduct $\Delta_0$.

Similarly in noncommutative case, the wavefunction lives in $\mathcal A_\theta \otimes \mathcal A_\theta$ and transforms according to the twisted coproduct $\Delta_\theta$.

For $\theta^{\mu \nu}=0$ we require that the physical wave functions describing identical particles are either symmetric (bosons) or antisymmetric (fermions).

That is we work with either the symmetrized or antisymmetrized tensor product
\begin{eqnarray}
\phi \otimes_{S,A} \chi &\equiv& \frac{1}{2}\left(\phi \otimes \chi
\pm\chi
\otimes \phi \right)
\end{eqnarray}

In a Lorentz-invariant theory, these relations have to hold in all frames of reference. The twisted coproduct action of the Lorentz group is not compatible with the usual symmetrization/antisymmetrization.

Let $\tau_0$ be the statistics (flip) operator associated with exchange for $\theta^{\mu \nu}=0$:
\begin{equation}
\tau_0(\phi \otimes \chi) = \chi \otimes \phi.
\end{equation}

For $\theta^{\mu \nu}=0$ , we have the axiom that $\tau_0$ is superselected. In particular, for Lorentz group action, $\Delta_{0}(\Lambda) = \Lambda \otimes \Lambda$, must and {\it does }commute with the statistics operator:
\begin {equation}
\tau_0\ \Delta_{0}(\Lambda)=\Delta_{0}(\Lambda) \tau_0.
\end{equation}

Given an element $\phi~\otimes~\chi$ of the tensor product, the physical Hilbert spaces can be constructed from the elements
\begin{equation}
\left(\frac{1 \pm \tau_0}{2}\right)~(\phi~\otimes~\chi).
\end{equation}
Now
\begin{equation}
\tau_{0} {\cal F}_{\theta} = {\cal F}_{\theta}^{-1} \tau_{0}
\end{equation}
so that
\begin{equation}
\tau_0\ \Delta_\theta(\Lambda) \neq\ \Delta_\theta(\Lambda) \tau_0.\nn
\end{equation}
It shows that the usual statistics is not compatible with the twisted coproduct.

But the new statistics operator \cite{statistics-uv-ir}
\begin{equation}
\tau_\theta~\equiv~{\cal F}_\theta^{-1}\tau_0 {\cal F}_\theta, \quad
\tau_\theta^2 = {\bf 1}\otimes {\bf 1}
\end{equation}
does commute with the twisted coproduct $\Delta_{\theta}$:
\begin{equation}
\Delta_\theta (\Lambda) = {\cal F}_\theta^{-1} \Lambda \otimes \Lambda~{\cal F}_\theta.
\end{equation}

The states constructed according to
\begin{equation}
\phi \otimes_{S_\theta} \chi \equiv
\left(\frac{1\,+ \tau_\theta}{2}\right)\,
(\phi\,\otimes\,\chi),
\end{equation}
\begin{equation}
\phi \otimes_{A_\theta} \chi \equiv \\
\left(\frac{1\,-
\tau_\theta}{2}\right)\,(\phi\,\otimes\,\chi)
\end{equation}
form the physical two-particle Hilbert spaces of (generalized) bosons and fermions and obey twisted statistics.
\section{The Pauli Principle}
In Ref. \cite{Chakraborty} the statistical potential $V_{\textrm{\tiny{STAT}}}$ between two identical fermions at inverse temperature $\beta$ has been computed:
\begin{equation}
{\textrm{exp}}\Big(-\beta V_{\textrm{\tiny{STAT}}} ({\overrightarrow{\bf x}}_{1}, {\overrightarrow{\bf x}}_{2})\Big) = \langle {\overrightarrow{\bf x}}_{1}, {\overrightarrow{\bf x}}_{2}|e^{-\beta H}|{\overrightarrow{\bf x}}_{1},{\overrightarrow{\bf x}}_{2}\rangle, \nonumber
\end{equation}
\begin{equation}
H = \frac{1}{2m}({\overrightarrow{\bf p}}_{1}^{2}+{\overrightarrow{\bf p}}_{2}^{2}).\nonumber
\end{equation}

Here $|{\overrightarrow{\bf x}}_{1},{\overrightarrow{\bf x}}_{2}\rangle$ has twisted antisymmetry:
\begin{equation}
\tau_{\theta}|{\overrightarrow{\bf x}}_{1},{\overrightarrow{\bf x}}_{2}\rangle = - |{\overrightarrow{\bf x}}_{1},{\overrightarrow{\bf x}}_{2}\rangle. \nonumber
\end{equation}

It is explicitly shown not to have an infinitely repulsive core, establishing the violation of Pauli principle, as we had earlier suggested.

This result has phenomenological consequences such as Pauli forbidden transitions (on which there are stringent limits). For example, in the Borexino and Super Kamiokande experiments, the forbidden transitions from $O^{16}$($C^{12}$) to $\tilde{O}^{16}$($\tilde{C}^{12}$) where the tilde nuclei have an extra nucleon in the filled $1S_{1/2}$ level are found to have lifetimes greater than $10^{27}$ years. There are also experiments on forbidden transitions to filled K-shells of crystals done in Maryland which give branching ratios less than $10^{-25}$ for such transitions. The consequences of these results to noncommutative models are yet to be studied.

\subsection{Bounds from NEMO experiment}

There have been experiments done by NEMO-2 ~\cite{Barabash}, where
searches for non-Paulian atoms and transitions to non-Paulian states
have been conducted. Data from these experiments may be used to
obtain a bound for the noncommutativity parameter $\theta$.

Non-Paulian atoms are those whose atomic orbitals are filled
violating the Pauli Exclusion Principle. As an example, non-Paulian
Carbon has as its atomic configuration, $1s^3 2s^2 2p^1$. The
presence of just a single electron in the outermost shell of
non-Paulian Carbon makes it behave chemically like Boron, whose
atomic configuration is $1s^2 2s^2 2p^1$. Thus searching for
non-Paulian Carbon atoms in samples of Boron, we can get the
concentration of the former in the latter. Bounds on these values
were found by the NEMO experiments.

We consider the tensor product of two-electron states, each of which
is an eigenvector of a hydrogen atom Hamiltonian. We then see that
the two-electron states we consider are eigenvectors of the
co-product of $H$:
\begin{equation}{\Delta(H)=H_1\otimes 1 + 1\otimes
H_2}\end{equation} where $H_1$ and $H_2$ are hydrogen atom
Hamiltonians.

Consider two such states, $|1s 1s\rangle_{\theta}$ and $|1s
2s\rangle_{\theta}$, the $\theta$ in the suffix indicates these
states have been deformed according to twisted statistics as we have
seen in section (4). We know the hydrogen atom is invariant under
rotations and time translations, which implies that the twist
element formed from the generators of rotation and time
translations, commutes with this Hamiltonian and so it also commutes
with the co-product of this Hamiltonian. Then, these states are
eigenvectors of the above Hamiltonian (26) with two different energy
eigenvalues namely $2E_1$ and $E_1+E_2$ respectively. This means
that they must be orthogonal. However, for the existence of
non-Paulian atoms we expect to observe such transitions. We will
indeed do so, if we include an interaction part in the above
Hamiltonian. So we consider the new Hamiltonian to be:
\begin{equation}{H=H_1\otimes 1 + 1\otimes H_2 +
H_{int}.}\end{equation}

We then consider matrix elements of the interaction,
$_{\theta}\langle 1s,1s|H_{int}|1s,2s\rangle_{\theta}$, and find
them to be non-zero. $H_{int}$ arises from the Coulomb interaction
between the electrons.

The results of this work will appear in a forthcoming paper
~\cite{ApbAjGmPp}.

\section{Cosmic Microwave Background (CMB)}
The COBE satellite, in 1992, detected anisotropies in the CMB radiation, which led to the conclusion that the early universe was not smooth: There were small perturbations in the photon-baryon fluid.

The perturbations could be due to the quantum fluctuations in the inflaton (the scalar field driving inflation). These fluctuations act as seeds for the primordial perturbations over the smooth universe. Thus according to these ideas, the early universe had inhomogeneities and we observe them today in the distribution of large scale structure and anisotropies in the CMB radiation.

The temperature field in the sky can be expanded in spherical harmonics:
\be
{\Delta T(\hat{n}) \over T} = \sum_{l m} a_{l m} Y_{l m}(\hat{n}).
\ee

The $a_{lm}$ can be written in terms of perturbations to Newtonian potential $\Phi$
\be
a_{lm} = 4\pi(-i)^l\int\frac{d^3k}{(2\pi)^3}~\Phi(k)\Delta^T_l(k)Y^\ast_{lm}(\hat{k})
\ee
where $\Delta^T_l(k)$ are called the {\it transfer functions}.

In a noncommutative spacetime the quantum corrections to the inflaton are modified. The angular correlation functions aquire nondiagonal elements indicating rotational symmetry breaking in the universe.

The noncommutative angular correlation takes the form \cite{cmbpaper}
\begin{eqnarray}
\langle a_{lm} a^{*}_{l'm'}\rangle_{_\theta} &=& 8 \pi^{2}\int d^{}k \sum_{l''=0, \; l'': \textrm{\tiny{even}}}^{\infty}  i^{l+l'}(-1)^{l+m} (2l''+1)  \; k^{2} \Delta_{l}(k)\Delta_{l'}(k) P_{\Phi} (k)\nn \\
&& \times i_{l''}(\theta kH) \sqrt{(2l+1)(2l'+1)} \left( \begin{array}{ccc}
l & l' & l'' \\
0 & 0 & 0 \end{array} \right)\left( \begin{array}{ccc}
l & l' & l'' \\
-m & m' & 0 \end{array} \right)
\end{eqnarray}
where $\vec{\theta}^{0}=\theta(0,~0,~1)$ and $P_{\Phi}(k)$ is the power spectrum for mode $k$
\be
P_{\Phi}(k) = \frac{16 \pi G}{9 \epsilon} \frac{H^{2}}{k^{3}}\Big|_{aH =k},
\ee
$H$ is the Hubble parameter, $a$ is the cosmological scale factor and $\epsilon$ is the slow-roll parameter.

The angular correlator $\langle a_{lm} a^{*}_{l'm'}\rangle_{_\theta}$ is $\theta$ dependent indicating a preferred direction. Correlation functions are not invariant under rotations. They are not Gaussian either. It clearly breaks the statistical isotropy of the CMB radiation.

On fitting data \cite{cmbpaper2}, one finds an upper bound for the length scale associated with spacetime noncommutativity,
\be
\sqrt{\theta} \lesssim 10^{-17} \textrm{cm}
\ee
or a lower bound for the energy scale, $E$
\be
E \gtrsim 10^{3} \textrm{GeV}.
\ee
\section{Causality, Lorentz invariance and CPT}
\subsection{Causality and Lorentz invariance}
The $S$-matrix of quantum theories constructed on the GM plane is not Lorentz invariant. The reson is nothing but loss of causality.

Let ${\cal H}_I$ be the interaction Hamiltonian density in the interaction representation of the quantum theory. The interaction representation $S$-matrix is
\begin{equation}
S = T \exp \left( -i \int d^4 x {\cal H}_I(x) \right).
\end{equation}

Bogoliubov and Shirkov \cite{Bogoliubov} and then Weinberg \cite{Weinberg} long ago deduced from causality (locality) and relativistic invariance that ${\cal H}_I$ must a local field:
\begin{equation}
[{\cal H}_I(x), {\cal H}_I(y)] = 0,~~~x \sim y.
\end{equation}

But noncommutative theories are nonlocal and violate this condition: this is the essential reason for Lorentz noninvariace.

The effect Lorentz noninvariace on scattering amplitudes is striking. They depend on total incident momentum $\overrightarrow{P}_{\textrm{inc}}$ through the term
\begin{equation}
\theta^{0i}P^{\textrm{inc}}_{i}.
\end{equation}

The effects of $\theta^{\mu \nu}$ disappear in the center-of-mass system, or more generally if
\be
\theta^{0i}P^{\textrm{inc}}_{i} = 0.
\ee
But otherwise there is dependence on $\theta_{0i}$.

The decay $Z^{0} \longrightarrow 2 \gamma $ is forbidden even with noncommutativity in the approach of Aschieri {\it et. al.} More generally, a massive particle of spin $j$ does not decay into two massless particles of same helicity if $j$ is odd.

\subsection{CPT}
The noncommutative $S$-matrix transforms under ${\bf CPT}$ in the following way \cite{cpt},
\ba
{S}^{^{M, G}}_{\theta} &=& \text{T}~\exp~\Big[-i\int d^{4}x~{\cal H}^{^{M, G}}_{I
0}(x)~e^{{1 \over 2}\overleftarrow{\partial}\wedge P}\Big] \nn \\
&&\rightarrow \text{T}~\exp~\Big[i\int d^{4}x~{\cal H}^{^{M, G}}_{I
0}(x)~e^{-{1 \over 2}\overleftarrow{\partial}\wedge P}\Big] = ({S}^{^{M, G}}_{-\theta})^{-1},
\ea
where ${\cal H}^{^{M, G}}_{I 0}$ is the matter-gauge interaction hamiltonian density for $\theta^{\mu \nu}=0$.

After performing the spatial integration we can reduce $e^{\frac{1}{2}\overleftarrow{\partial}\wedge P}$ in the $S$-matrix to $e^{\frac{1}{2}\overleftarrow{\partial}_{0} \theta^{0i} P_{i}}$. Thus the effect of P and CPT is to reverse the sign of $\theta^{0i}$:
\begin{center}
P or CPT :~~$\theta^{0i} \rightarrow -\theta^{0i}$.
\end{center}
The $\theta^{0i}$ contributes to P, and more strikingly, to CPT violation.

The particle-antiparticle life times can differ to order $\theta^{0i}$:
\be
\tau_{\rm{\tiny{particle}}} - \tau_{\rm{\tiny{antiparticle}}} \cong  \theta^{0i} P^{\textrm{inc}}_{i}.
\ee

It can give rise to interesting effects such as mass difference in $K^0 - \bar{K}^0$ system and the $(g-2)$ difference of $\mu^+$-$\mu^-$. (See \cite{Joseph:2008fz} for bounds on $\theta$ estimated from these effects.)

\section{Conclusions}
Spacetime noncommutativity deforms statistics and so generically violate causality in noncommutative quantum theories. Such violations lead to many interesting features such as $(i.)$ modification of Pauli principle causing forbidden atomic transitions, $(ii.)$ correlations of observables in spacelike regions giving rise to anisotropies in the CMB radiation, $(iii.)$ Lorentz and CPT violations in scattering amplitudes.

It is shown that there are specific predictions that may be observable. Bounds on noncommutativity parameter are given in the context of different experimental measurements.

\begin{acknowledgements}
The authors acknowledge their collaboration with Earnest Akofor, T.
R. Govindarajan, Sang Jo, Gianpiero Mangano, Aleksandr Pinzul,
Amilcar Queiroz, Babar Qureshi, P. Teotonio-Sobrinho and Sachin
Vaidya. APB would like to thank the organizers of the Spinstat2008
workshop for their kind hospitality he enjoyed at Trieste. APB also
thanks Alberto Ibort and the Universidad Carlos III de Madrid for
their kind hospitality and support. This work was partially
supported by the US Department of Energy under grant number
DE-FG02-85ER40231.
\end{acknowledgements}

\end{document}